# Scattering state of Klein-Gordon Particles by $q$-parameter hyperbolic Poschl-Teller potential

AkpanNdemIkot[1*],Hillary P.Obong[1],Israel O.Owate[1],Michael C.Onyeaju[1]and Hassan Hassanabadi[2]

[1] Theoretical Physics group,Department of Physics, University of Port Harcourt, P M B 5323 Choba ,Port Harcourt -Nigeria.

[2]Physics Department, University of Shahrood, Shahrood, Iran.

.

## Abstract

The one-dimensional Klein-Gordon equation for equal vector and scalar q-parameter hyperbolic Poschl-Teller potential is solved in terms of the hypergeometric functions. We calculate in details the solutions of the scattering and bound states. By virtue of the conditions of equation of continuity of the wave functions, we obtained explicit expressions for the reflection and transmission coefficients and energy equation for the bound state solutions.

*Corresponding author :ndemikotphysics@gmail.com

# 1 Introduction

The scattering states of the relativistic and non-relativistic wave equation in recent times have received a great attention in physics [1-9]. Scattering and bound state solutions of asymmetric Hulthen potential have been obtained by Arda et al [10]. Also, the bound state and scattering state of Klein-Gordon equation with effective mass formalism have been studied by Arda and Sever[11]. Aydogdu et al [12] examined the scattering and bound states of massive Dirac equation with Wood-Saxon potential. Villalba and Rojas [13] in their investigation had shown the relation between the bound state energy eigenvalues and transmission resonances for the Klein-Gordon particle with Woods-Saxon potential as the same for the Dirac particle.Hassanabadi et al[14-15] investigated the scattering state of relativistic spinless particles with Kink-like potential and scattering of Woods-Saxon potential in the framework of minimal length. The scattering states ofrelativistic and non-relativistic particles with Woods-Saxon potential and mass dependent Woods-Saxon have been studied by Alpdogan etal[16] and Arda et al[17]. Recently, Alpdogan studied the scattering of massive Dirac particles with generalized asymmetric Woods-Saxon potential[18]. Yanar et al studied the scattering and bound state of Duffin-Kemmer-Petiau particles with q-parametr hyperbolic Poschl-Teller potential(qHPT)[19] .The qHPT is defined as follows

$$V(x) = \frac{4\lambda(\lambda-1)}{\cosh_q^2(\alpha x)}$$

$$= 4\lambda(\lambda-1)\left[\theta(-x)\frac{e^{2\alpha x}}{\left(1+qe^{2\alpha x}\right)^2} + \theta(x)\frac{e^{-2\alpha x}}{\left(1+qe^{-2\alpha x}\right)^2}\right] \qquad (1)$$

where $\theta(x)$ is the step function, $q$ is the deformation parameter and $q \neq 0$, $\lambda$ is the height of the potential , $\lambda \neq 0, \lambda \neq 1$ and $\alpha$ is the range of the potential barrier. This potential plays important role in describing the interactions in molecular, atomic and nuclear physics. One of the findings of physics is to understand the structures of nucleus, atoms, molecules and other material object. Thus, the sole aim of physicists is to create unique models that contain the physically motivated potential that will describe the interactions between particles. A few numbers of these potential models have been identified to describe the interaction in the nuclei and diatomic and polyatomic molecules [20-21]. The applications of the Poschl-Teller-like type potential for analyzing the bound states energy of $\Lambda$-particle in hypernuclei in nuclear physics had been reported [19]. Furthermore, the (qHPT) may found much usefulness in molecular,atomic and nuclear physics.

Understanding of our knowledge about fine scale systems have been gained by investigations scattering and bounded state of such a systems.Therefore, the scattering problem has become an interesting topic in relativistic or non-relativisticquantum mechanics. In the case of non-relativistic scattering problem, it has been shown that transmission and reflection coefficients take 1 and 0, respectively, as external potential has well-behaved at infinity for the zero energy limit [22-24]. However, reflection coefficient goes to zero while transmission coefficient goes to unity in the zero energy limit when external potential supports a half-bound state. This situation was called as transmission resonance [25] which is escorted by fluctuations phenomena. The transmission resonance concept has been recently generalized to the relativistic case [26].

In this work, we attempt to study the solution of scattering state of the Klein Gordon equation with equal vector and scalar (qHPT) potential. Our aim will be to calculate in details the reflection(R) and transmission (T) coefficients and obtained the bound state solution of the qHPT using the equation of continuity of the wave function.

## 2 Scattering state solutions of Klein-Gordon equation for the qHPT

The time-independent Klein-Gordon equation with equal scalar $S(x)$ and vector $V(x)$ potentials can be written as[27],

$$\frac{d^2\psi(x)}{dx^2} + \left\{E^2 - m^2 - 2(E+m)V(x)\right\}\psi(x) = 0, \ (\hbar = c = 1), \quad (2)$$

Where E is the relativistic energy of the particles and m is the mass of the particle. In order to obtain the scattering solution for qHPT, we consider both $x < 0$ and $x > 0$ at $|E| > m$. We intend to study the scattering of equation (2),we now seek the wave function for the case $x < 0$. By substituting Eq.(1) into Eq. (2), we have

$$\frac{d^2\psi(x)}{dx^2} + \left\{E^2 - m^2 - \frac{8\lambda(\lambda-1)(E+m)e^{2\alpha x}}{\left(1+qe^{2\alpha x}\right)^2}\right\}\psi(x) = 0 \quad (3)$$

To solve equation (3), we used a new variable defined as $z = -qe^{2\alpha x}$ which yields

$$z(1-z)\frac{d^2\psi}{dz^2} + (1-z)\frac{d\psi}{dz} + \frac{1}{z(1-z)}\left\{\gamma_1 z^2 + \gamma_2 z + \gamma\right\}\psi(z) = 0 \quad (4)$$

Where,

$$\gamma_1 = \frac{1}{4\alpha^2}\left(E^2 - m^2\right), \qquad (5)$$

$$\gamma_2 = \frac{2}{\alpha^2}\left(\frac{(E+m)\lambda(\lambda-1)}{q} + \frac{m^2 - E^2}{4}\right) \qquad (6)$$

$$\gamma_3 = \frac{1}{4\alpha^2}\left(E^2 - m^2\right) \qquad (7)$$

Defining the wave function in Eq.(4) as $\psi(z) = z^\mu(1-z)^\nu\varphi(z)$, then Eq.(4) turns into the hypergeometric differential equation[28],

$$z(1-z)\frac{d^2\varphi}{dz^2} + \left[1 + 2\mu - \left(1 + 2\mu + 2\nu\right)z\right]\frac{d\varphi}{dz} - \left(\mu + \nu + \delta\right)\left(\mu + \nu - \delta\right)\varphi(z) = 0 \quad (7)$$

Where $\mu, \nu$ and $\delta$ are defined as follows,

$$\mu = \frac{ik}{2\alpha}, \nu = \frac{1}{2} \pm \frac{1}{2}\sqrt{1 - \frac{8(E+m)\lambda(\lambda-1)}{\alpha^2 q}},$$

$$\delta = \frac{ik}{2\alpha}, k^2 = \left(E^2 - m^2\right) \qquad (8)$$

Regarding properties of Hypergeometric functions the minus sign should be chosen for $\nu$. The solutions of equation (7) can be written in terms of the hypergeometric function as follows

$$\varphi(z<0) = A\,_2F_1\left(\mu+\nu+\delta, \mu+\nu-\delta, 1+2\mu, z\right) + B_2 z^{-2\mu}\,_1F_1\left(-\mu+\nu+\delta, -\mu+\nu-\delta, 1-2\mu, z\right)$$

$$(9)$$

The general solutions for $x < 0$, is given as

$$\psi(z<0) = Az^\mu\left(1-z\right)^\nu\,_2F_1\left(\mu+\nu+\delta, \mu+\nu-\delta, 1+2\mu, z\right)$$
$$+ Bz^{-\mu}\left(1-z\right)^\nu\,_2F_1\left(-\mu+\nu+\delta, -\mu+\nu-\delta, 1-2\mu, z\right) \qquad (10)$$

Next, we investigate the solution for $x > 0$. Again substituting Eq.(1) into Eq.(2) yield,

$$\frac{d^2\psi(x)}{dx^2} + \left\{E^2 - m^2 - \frac{8\lambda(\lambda-1)\left(E+m\right)e^{-2\alpha x}}{\left(1 + qe^{-2\alpha x}\right)^2}\right\}\psi(x) = 0 \qquad (11)$$

Defining the new variable $\tilde{z} = -qe^{-2\alpha x}$ and redefining the ansatz for the wave function as $\psi(\tilde{z}) = \tilde{z}^{\tilde{\mu}}(1-\tilde{z})^{\tilde{\nu}}f(\tilde{z})$, the solutions of equation (11) becomes,

$$\psi(\tilde{z}>0) = C\tilde{z}^{\tilde{\mu}}\left(1-\tilde{z}\right)^{\tilde{\nu}}\,_2F_1\left(\tilde{\mu}+\tilde{\nu}+\tilde{\delta}, \tilde{\mu}+\tilde{\nu}-\tilde{\delta}, 1+2\tilde{\mu}, \tilde{z}\right)$$
$$+ D\tilde{z}^{-\tilde{\mu}}\left(1-\tilde{z}\right)^{\tilde{\nu}}\,_2F_1\left(-\tilde{\mu}+\tilde{\nu}+\tilde{\delta}, -\tilde{\mu}+\tilde{\nu}-\tilde{\delta}, 1-2\tilde{\mu}, \tilde{z}\right) \qquad (12)$$

Where,

$$\tilde{\mu}=\frac{ik}{2\alpha},\tilde{v}=\frac{1}{2}\pm\frac{1}{2}\sqrt{1-\frac{8(E+m)\lambda(\lambda-1)}{q\alpha^2}},$$

$$\tilde{\delta}=\frac{ik}{2\hbar},k^2=\left(E^2-m^2\right) \qquad (13)$$

Here we also should choose the minus sign for $\tilde{v}$. We seek for the physical results of the problem under investigation, therefore in order to get this physical result, the solutions obtained have to be used with appropriate boundary conditions as $x\to-\infty$ and $x\to+\infty$. Applying the asymptotic solution to Eq.(10) for $x\to-\infty$, as $z_L\to0$ and $\left(1-z\right)^{v}\to1$, becomes,

$$\psi_L\left(x\to-\infty\right)=A\left(-\frac{1}{q}\right)^{\frac{ik}{2\alpha}}e^{ik}+B\left(-\frac{1}{q}\right)^{\frac{ik}{2\alpha}}e^{-ik} \qquad (14)$$

In order to find a plane wave traveling from left to right, we set $C=0$, in equation (12) and the asymptotic behavior of the right solution becomes,

$$\psi_R(x\to\infty)=D\left(-q\right)^{\frac{ik}{2\alpha\hbar c}}e^{\frac{ik}{\hbar c}} \qquad (15)$$

Now in order to give an explicit expressions for the coefficients, we use the continuity conditions of the wave function and its derivative defined as,

$$\psi_L(x=0)=\psi_R(x=0),$$
$$\psi_L'(x=0)=\psi_R'(x=0) \qquad (16)$$

Where the prime denote differential with respect to $x$. Applying these conditions on the wave function and matching the wave function at $x=0$, we get

$$AQ_1U_1+BQ_2U_2=DQ_3U_3 \qquad (17)$$

$$A\left(Q_4U_4+Q_5U_5+Q_6S_1U_6\right)+B\left(Q_7U_7+Q_8U_8+Q_9S_2U_9\right)=D\left(Q_{10}U_{10}+Q_{11}U_{11}+Q_{12}S_3U_{12}\right) \qquad (18)$$

Where the following abbreviations have been used

$$Q_1=\left(-q\right)^{\mu}\left(1+q\right)^{v},Q_2=\left(-q\right)^{-\mu}\left(1+q\right)^{v},Q_3=\left(-q\right)^{-\tilde{\mu}}\left(1+q\right)^{\tilde{v}}$$

$$Q_4=-2\alpha\mu\left(-q\right)^{\mu}\left(1+q\right)^{v},Q_5=2\alpha v\left(-q\right)^{\mu+1}\left(1+q\right)^{v-1},Q_6=-2\alpha\left(-q\right)^{\mu+1}\left(1+q\right)^{v},$$

$$Q_7=2\alpha\mu\left(-q\right)^{-\mu}\left(1+q\right)^{v},Q_8=2\alpha v\left(-q\right)^{-\mu+1}\left(1+q\right)^{v-1},Q_9=-2\alpha\left(-q\right)^{-\mu+1}\left(1+q\right)^{v}$$

$$Q_{10}=-2\alpha\tilde{\mu}\left(-q\right)^{-\tilde{\mu}}\left(1+q\right)^{\tilde{v}},Q_{11}=-2\alpha\tilde{v}\left(-q\right)^{-\tilde{\mu}+1}\left(1+q\right)^{\tilde{v}-1},Q_{12}=2\alpha\left(-q\right)^{-\tilde{\mu}+1}\left(1+q\right)^{\tilde{v}}$$

$$S_1=\frac{\left(\mu+v+\delta\right)\left(\mu+v-\delta\right)}{\left(1+2\mu\right)},S_2=\frac{\left(-\mu+v+\delta\right)\left(-\mu+v-\delta\right)}{\left(1-2\mu\right)},S_3=\frac{\left(-\tilde{\mu}+v+\delta\right)\left(-\tilde{\mu}+v-\delta\right)}{\left(1-2\tilde{\mu}\right)} \qquad (19)$$

and

$U_1 = {}_2F_1\left(\mu+\nu+\delta,\mu+\nu-\delta,1+2\mu,-q\right), U_2 = {}_2F_1\left(-\mu+\nu+\delta,-\mu+\nu-\delta,1-2\mu,-q\right),$

$U_3 = {}_2F_1\left(-\tilde{\mu}+\tilde{\nu}+\tilde{\delta},-\tilde{\mu}+\tilde{\nu}-\tilde{\delta},1-2\tilde{\mu},-q\right), U_4 = {}_2F_1\left(\mu+\nu+\delta,\mu+\nu-\delta,1+2\mu,-q\right),$

$U_5 = {}_2F_1\left(\mu+\nu+\delta,\mu+\nu-\delta;1+2\mu,-q\right), U_6 = {}_2F_1\left(\mu+\nu+\delta+1,\mu+\nu-\delta+1;2+2\mu,-q\right),$

$U_7 = {}_2F_1\left(-\mu+\nu+\delta,-\mu+\nu-\delta,1-2\mu,-q\right), U_8 = {}_2F_1\left(-\mu+\nu+\delta,-\mu+\nu-\delta,1-2\mu,-q\right),$

$U_9 = {}_2F_1\left(-\mu+\nu+\delta+1,-\mu+\nu-\delta+1,2-2\mu,-q\right), U_{10} = {}_2F_1\left(-\tilde{\mu}+\tilde{\nu}+\tilde{\delta},-\tilde{\mu}+\tilde{\nu}-\tilde{\delta},1-2\tilde{\mu},-q\right),$

$U_{11} = {}_2F_1\left(-\tilde{\mu}+\tilde{\nu}+\tilde{\delta},-\tilde{\mu}+\tilde{\nu}-\tilde{\delta},1-2\tilde{\mu},-q\right), U_{12} = {}_2F_1\left(-\tilde{\mu}+\tilde{\nu}+\tilde{\delta}+1,-\tilde{\mu}+\tilde{\nu}-\tilde{\delta}+1,2-2\tilde{\mu},-q\right)$ (20)

And the property of hypergeometric function $\dfrac{d}{ds}{}_2F_1\left(a,b,c,s\right) = \dfrac{ab}{c}{}_2F_1\left(a+1,b+1,c+1,s\right)$

[29] has been used in obtaining Eq.(18).The one dimensional current density for the Klein-Gordon equation in terms of the relativistic units($\hbar = c = 1$) is given by[30]

$$J(x) = \frac{i}{2m}\left(\psi^*\nabla_x\psi - \psi\nabla_x\psi^*\right) \qquad (21)$$

The reflection and transmission coefficients are defined in terms of current density as follows

$$R = \frac{j_{ref}}{j_{inc}} = \left|\frac{B}{A}\right|^2,$$

$$T = \frac{j_{trans}}{j_{inc}} = \left|\frac{D}{A}\right|^2 \qquad (22)$$

Now using these equations and after a tedious algebra, we obtain the reflection and transmission coefficients as,

$$R = \left|\frac{B}{A}\right|^2 = \left|\frac{Q_4U_4 + Q_5U_5 + Q_6S_1U_6 - \dfrac{Q_1U_1}{Q_3U_3}\left(Q_{10}U_{10} + Q_{11}U_{11} + Q_{12}S_3U_{12}\right)}{\dfrac{Q_2U_2}{Q_3U_3}\left(U_{10}Q_{10} + Q_{11}U_{11} + Q_{12}S_3U_{12}\right) - \left(Q_7U_7 + Q_8U_8 + Q_9S_2U_9\right)}\right|^2 \qquad (21)$$

$$T = \left|\frac{D}{A}\right|^2 = \left|\frac{\dfrac{Q_2U_2}{Q_3U_3}\left(Q_4U_4 + Q_5U_5 + Q_6S_1U_6\right) - \dfrac{Q_1U_1}{Q_3U_3}\left(Q_7U_7 + Q_8U_8 + Q_9S_2U_9\right)}{\dfrac{Q_2U_2}{Q_3U_3}\left(U_{10}Q_{10} + Q_{11}U_{11} + Q_{12}S_3U_{12}\right) - \left(Q_7U_7 + Q_8U_8 + Q_9S_2U_9\right)}\right|^2 \qquad (22)$$

<span style="color:red">We have depicted in Fig. 5 that sum of reflection and transmission coefficient approaches to unite.</span>

## 3 Bound state solutions of Klein-Gordon Equation for the qHPT

In order to find the bound state solutions for the Klein-Gordon particle with qHPT, we map $8\lambda(\lambda-1) \rightarrow -V_0$.

### A. Bound state solutions in the negative region ($x < 0$)

The bound state solutions can be found in this region by changing the variable $z = -qe^{2\alpha x}$ and taking into consideration $8\lambda(\lambda-1) \rightarrow -V_0$ and Eq.(3) becomes

$$z(1-z)\frac{d^2\psi}{dz^2} + (1-z)\frac{d\psi}{dz} + \frac{1}{z(1-z)}\left\{\beta_1 z^2 + \beta_2 z + \beta\right\}\psi(z) = 0 \quad (23)$$

Where,

$$\beta_1 = \frac{1}{4\alpha^2}\left(E^2 - m^2\right), \quad (24)$$

$$\beta_2 = -\frac{1}{2\alpha^2}\left(\frac{V_0(E+m)}{q} + (E^2 - m^2)\right) \quad (25)$$

$$\beta_3 = \frac{1}{4\alpha^2}\left(E^2 - m^2\right) \quad (26)$$

Taking a wave function of the form, $\psi(z) = z^{\mu_1}(1-z)^{\nu_1}\varphi(z)$, then Eq.(23) turns into the hypergeometric differential equation[28],

$$z(1-z)\frac{d^2\varphi}{dz^2} + \left[1 + 2\mu_1 - (1 + 2\mu_1 + 2\nu_1)z\right]\frac{d\varphi}{dz} - (\mu_1 + \nu_1 + \delta_1)(\mu_1 + \nu_1 - \delta_1)\varphi(z) = 0 \quad (27)$$

with $\mu_1, \nu_1$ and $\delta_1$ are defining as follows,

$$\mu_1 = \frac{ik}{2\alpha}, \nu_1 = \frac{1}{2} \pm \frac{1}{2}\sqrt{1 + \frac{(E+m)V_0}{\alpha^2 q}},$$

$$\delta_1 = \frac{ik}{2\alpha}, k^2 = \left(E^2 - m^2\right) \quad (28)$$

That for $\nu_1$ the correct sign is minus. The general solutions for $x < 0$, is given as

$$\psi(z < 0) = A_1 z^{\mu_1}(1-z)^{\nu_1} \, _2F_1\left(\mu_1 + \nu_1 + \delta_1, \mu_1 + \nu_1 - \delta_1, 1 + 2\mu_1, z\right)$$
$$+ B_1 z^{-\mu_1}(1-z)^{\nu_1} \, _2F_1\left(-\mu_1 + \nu_1 + \delta_1, -\mu_1 + \nu_1 - \delta_1, 1 - 2\mu_1, z\right) \quad (29)$$

### B. Bound state solutions in the positive region ($x > 0$)

In the positive region, we defined the variable, $\tilde{z} = -qe^{-2\alpha x}$ with $8\lambda(\lambda-1) \rightarrow -V_0$ and taking the wave function $\psi(\tilde{z}) = \tilde{z}^{\mu_1}\left(1-\tilde{z}\right)^{\nu_1}\varphi(\tilde{z})$ and following the same procedures as the case of negative region, we obtain the solutions for the positive region as follows:

$$\psi(\tilde{z} > 0) = C_1 \tilde{z}^{\tilde{\mu}_1} \left(1 - \tilde{z}\right)^{\tilde{\nu}_1} {}_2F_1\left(\tilde{\mu}_1 + \tilde{\nu}_1 + \tilde{\delta}_1, \tilde{\mu}_1 + \tilde{\nu}_1 - \tilde{\delta}_1, 1 + 2\tilde{\mu}_1, \tilde{z}\right)$$
$$+ D_1 z^{-\tilde{\mu}_1} \left(1 - \tilde{z}\right)^{\tilde{\nu}_1} {}_2F_1\left(-\tilde{\mu}_1 + \tilde{\nu}_1 + \tilde{\delta}_1, -\tilde{\mu}_1 + \tilde{\nu}_1 - \tilde{\delta}_1, 1 - 2\tilde{\mu}_1, \tilde{z}\right) \qquad (30)$$

Where,

$$\tilde{\mu}_1 = \frac{ik}{2\alpha}, \tilde{\nu}_1 = \frac{1}{2} \pm \frac{1}{2}\sqrt{1 + \frac{2(E + mc^2)V_0}{\alpha^2 q}},$$

$$\tilde{\delta}_1 = \frac{ik}{2\alpha}, k^2 = \left(E^2 - m^2\right) \qquad (31)$$

And as mention in above pages the minus sing should be chosen. In order to find the equation for the energy eigenvalues, we set $B = D = 0$ and used the conditions of continuity for the wave function as, $\psi_L(x = 0) = \psi_R(x = 0), \psi'_L(x = 0) = \psi'_R(x = 0)$ and we get

$$A_1 a_1 K_1 = C_1 b_1 K_2 \qquad (32)$$

$$A_1\left(a_2 K_3 + a_3 K_4 + a_5 K_5\right) = C_1\left(b_2 K_6 + b_3 K_7 + b_4 K_8\right) \qquad (33)$$

Where,

$$a_1 = (-q)^{\mu_1}(1 + q)^{\nu_1}, K_1 = {}_2F_1\left(\mu_1 + \nu_1 + \delta_1, \mu_1 + \nu_1 - \delta_1, 1 + 2\mu_1; -q\right), b_1 = (-q)^{\tilde{\mu}_1}(1 + q)^{\tilde{\nu}_1}$$

$$K_2 = {}_2F_1\left(\tilde{\mu}_1 + \tilde{\nu}_1 + \tilde{\delta}_1, \tilde{\mu}_1 + \tilde{\nu}_1 - \tilde{\delta}_1, 1 + 2\tilde{\mu}_1; -q\right), a_2 = 2\alpha\mu_1(-q)^{\mu_1}(1 + q)^{\nu_1}, a_3 = 2\alpha\nu(-q)^{\mu_1 + 1}(1 + q)^{\nu_1 - 1}$$

$$a_4 = 2\alpha(-q)^{\mu_1 + 1}(1 + q)^{\nu_1}, K_3 = {}_2F_1\left(\mu_1 + \nu_1 + \delta_1, \mu_1 + \nu_1 - \delta_1, 1 + 2\mu_1; -q\right),$$

$$K_4 = {}_2F_1\left(\mu_1 + \nu_1 + \delta_1, \mu_1 + \nu_1 - \delta_1, 1 + 2\mu_1; -q\right), b_2 = -2\alpha\tilde{\mu}_1(-q)^{\tilde{\mu}_1}(1 + q)^{\tilde{\nu}_1}, b_3 = 2\alpha\tilde{\nu}_1(-q)^{\tilde{\mu}_1 + 1}(1 + q)^{\tilde{\nu}_1 - 1},$$

$$b_4 = -2\alpha(-q)^{\tilde{\mu}_1 + 1}(1 + q)^{\tilde{\nu}_1}, K_5 = \frac{(\mu_1 + \nu_1 + \delta_1)(\mu_1 + \nu_1 - \delta_1)}{1 + 2\mu_1} {}_2F_1\left(\mu_1 + \nu_1 + \delta_1 + 1, \mu_1 + \nu_1 - \delta_1 + 1, 2 + 2\mu_1; -q\right)$$

$$K_6 = {}_2F_1\left(\tilde{\mu}_1 + \tilde{\nu}_1 + \tilde{\delta}_1, \tilde{\mu}_1 + \tilde{\nu}_1 - \tilde{\delta}_1, 1 + 2\tilde{\mu}_1; -q\right), K_7 = {}_2F_1\left(\tilde{\mu}_1 + \tilde{\nu}_1 + \tilde{\delta}_1, \tilde{\mu}_1 + \tilde{\nu}_1 - \tilde{\delta}_1, 1 + 2\tilde{\mu}_1; -q\right),$$

$$K_8 = \frac{(\tilde{\mu}_1 + \tilde{\nu}_1 + \tilde{\delta}_1)(\tilde{\mu}_1 + \tilde{\nu}_1 - \tilde{\delta}_1)}{1 + 2\tilde{\mu}_1} {}_2F_1\left(\tilde{\mu}_1 + \tilde{\nu}_1 + \tilde{\delta}_1 + 1, \tilde{\mu}_1 + \tilde{\nu}_1 - \tilde{\delta}_1 + 1, 2 + 2\tilde{\mu}_1; -q\right) \qquad (34)$$

Equation (32) and (33) admit a solution if and only if its determinant is zero [31]. This thus provides a condition for getting the energy eigenvalues as

$$b_1 a_2 K_2 K_3 + b_1 a_3 K_2 K_4 + b_1 a_4 K_2 K_5 = a_1 b_2 K_1 K_6 + a_1 b_3 K_1 K_7 + a_1 b_4 K_1 K_8 \qquad (35)$$

The energy equation (35) is a complicated transcendental equation and can only be solved numerically.

## 4 Results and Discussion

The behaviours of the qHPTpotential as a function of x are displayed in Fig.1-4 for various values of qHPT potential parameters. For $q = 0$ and $8\lambda(\lambda - 1) \rightarrow V_0$, the qHPT potential turns to Cusp potential which is consistent one reported in ref.[32].However, if we map $E - M = \frac{2\mu}{\hbar^2}$ and $E + M = E_{nl}$ in Eq.(35) our results reduced to the cusp potential for the non-relativistic regime as reported in Ref.[33].

**5 Conclusions**

We solved the relativistic spin-less Klein-Gordon particles for the qHPT potential and calculate the wave functions that describe the scattering states in terms of hypergeometric function. By mapping of the potential parameter of qHPT potential, we also solved the bound state solution of the Klein –Gordon equation. By virtue of the equation of continuity of the wave function and the asymptotic properties of the solutions of the wave function, we calculate in details the reflection, transmission coefficients and bound state solution by the vanishes of the determinant of the coefficients of the wave functions for the qHPT potential .This study can find many applications of physics especially in the interaction of nuclei in nuclear physics.

**Acknowledgment**

It's our pleasure that we have faced to right and faithful comments of referees to a have better article.

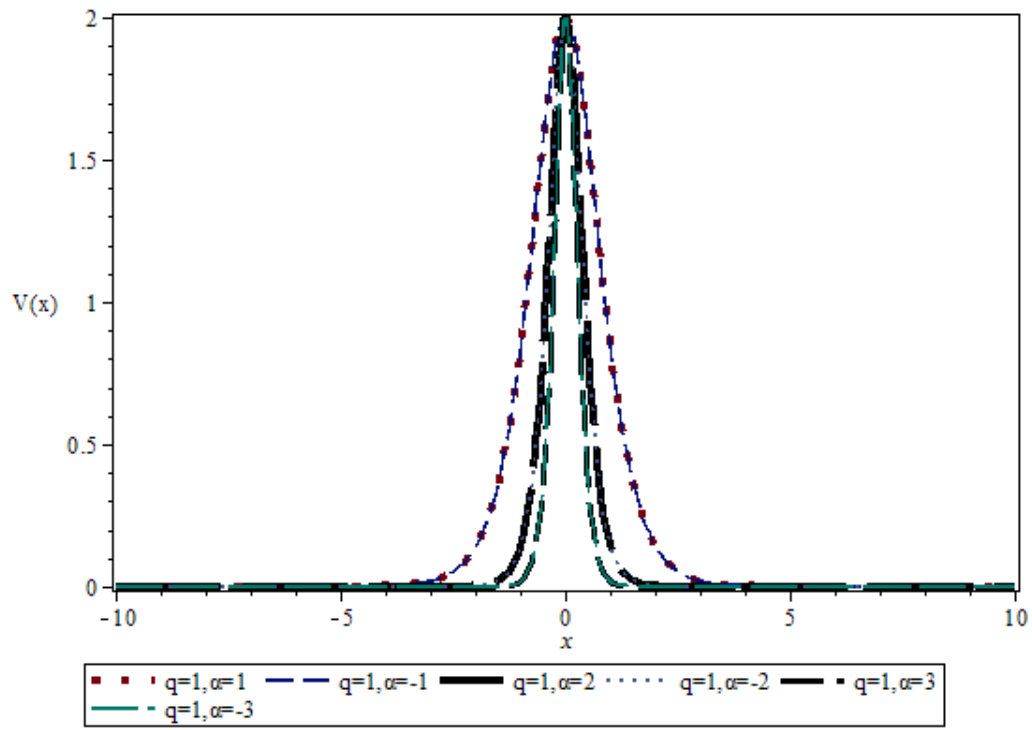

Fig.1: The plot of the qHPTpotential as a function of position with $q = 1, \alpha = 1, 2$ and $3$ for $x < 0$ and $x > 0$ respectively.

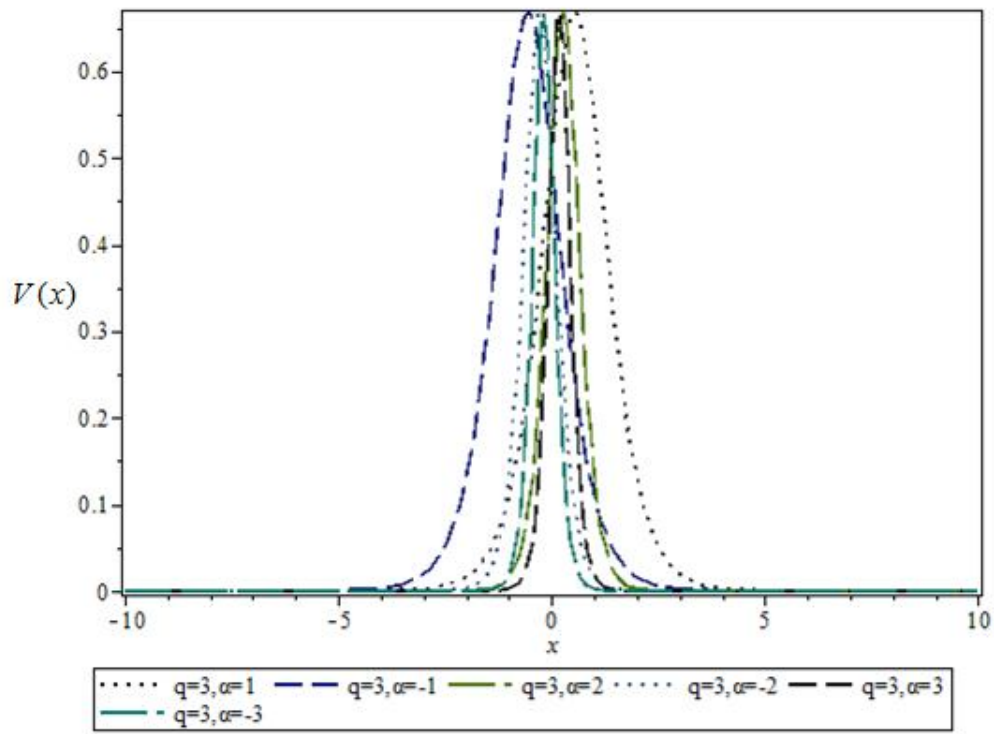

Fig.2: The plot of the qHPT potential as a function of position with $q = 3, \alpha = 1, 2$ and 3 for $x < 0$ and $x > 0$ respectively.

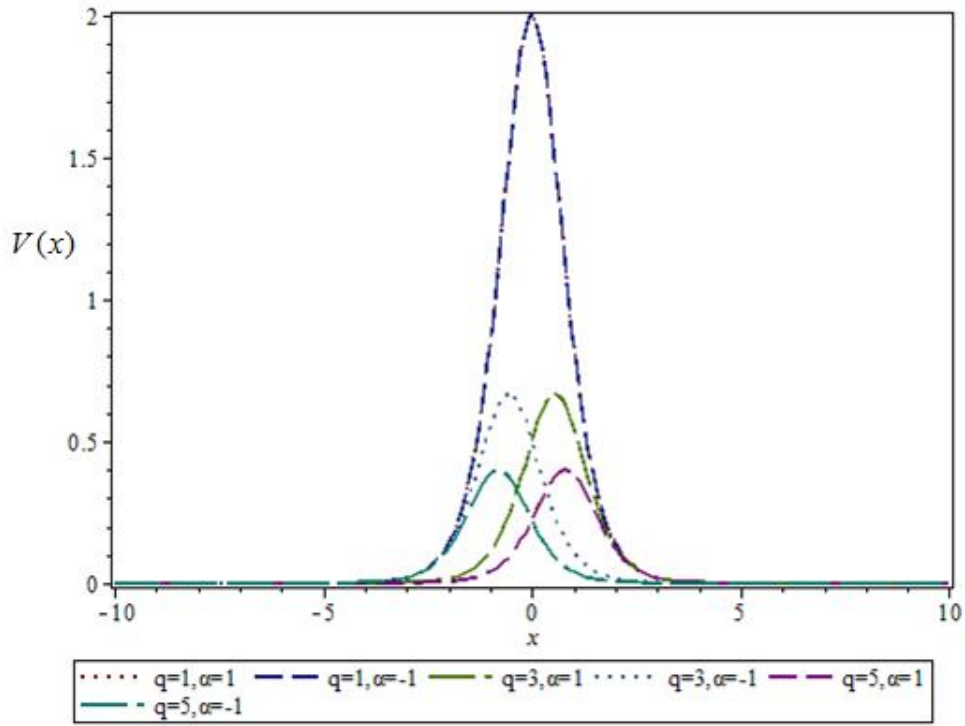

Fig.3: The plot of the qHPT potential as a function of position with $q = 1, 3$ and $5$ for $x < 0$ and $x > 0$ respectively.

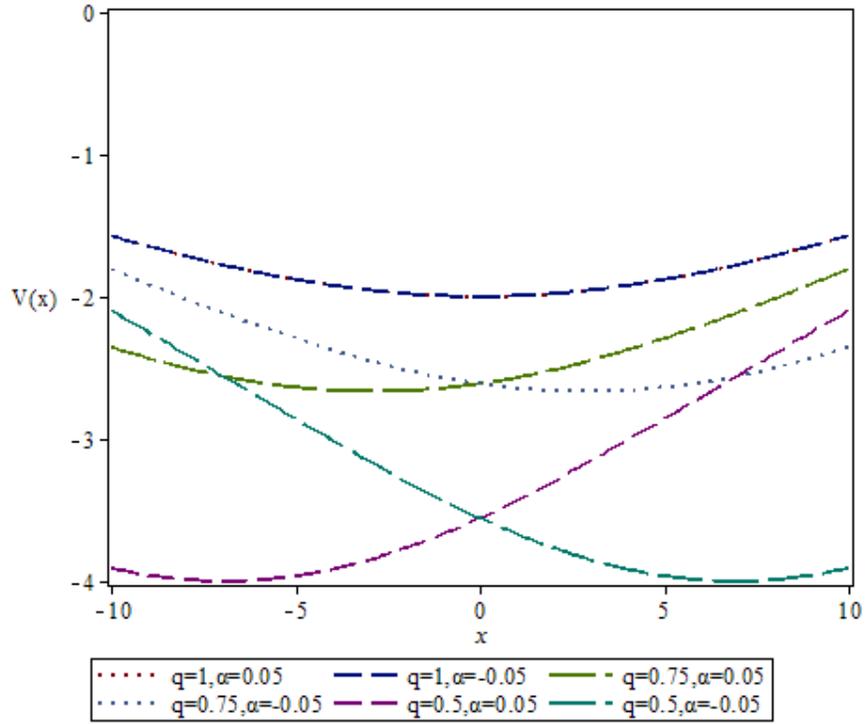

Fig.4: The plot of the qHPT potential as a function of position $x$ with $q = 1, 0.75$ and $0.5$ and $\alpha = 0.05$ for $x < 0$ and $x > 0$ respectively

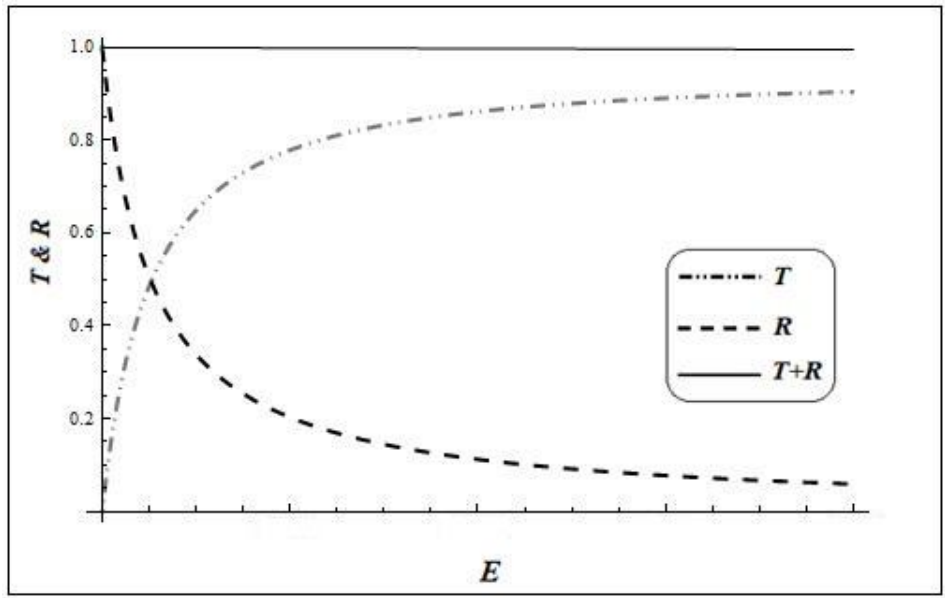

Fig. 5:Transmission ($T$) and Reflection ($R$) coefficients vs. $E$ is depicted.